\documentclass[aps,prl,twocolumn,superscriptaddress]{revtex4}
\usepackage[intlimits]{amsmath}
\usepackage{amsfonts}
\usepackage{graphics}
\usepackage{subfigure}
\usepackage[usenames]{color}
\usepackage{epstopdf,epsfig,todonotes}
\usepackage{xcolor}
\usepackage{indentfirst}
\usepackage{enumerate}
\usepackage{tikz}
\usepackage{mathtools}

\usepackage{graphicx}
\usepackage{amsmath}
\usepackage{amssymb, mathtools, mathrsfs}
\usepackage{hyperref}
\DeclareMathOperator{\res}{Res}
\newcommand\be            {\begin{equation}}
\newcommand\ee            {\end{equation}}
\def\d{{\rm d}}
\def \ov{\overline}

\def\bea{\begin{eqnarray}}
\def\eea{\end{eqnarray}}
\newcommand{\C}{\mathbb{C}}

\newcommand\EN           {\end{equation}}
\newcommand\bes           {\begin{subequations}}
\newcommand\esu           {\end{subequations}}

\newcommand\D{\mathcal{D}}

\newcommand{\I}{\mathcal{I}}

\def\3pt#1#2#3{{\langle{#1}\vert{#2}\vert{#3}\rangle}}

\begin{document}

\title{ Soliton shielding of the focusing nonlinear Schr\"odinger equation}

\author{Marco Bertola}
\affiliation{SISSA, via Bonomea 265, 34136, Trieste, Italy}
\affiliation{Concordia University, 1455 av. de Maisonneuve W.,  Montr\'eal Canada}
\affiliation{INFN, Sezione di Trieste}
\author{Tamara Grava}
\affiliation{SISSA, via Bonomea 265, 34136, Trieste, Italy}
\affiliation{School of Mathematics, University of Bristol, Fry Building, Bristol,
BS8 1UG, UK}
\affiliation{INFN, Sezione di Trieste}
\author{Giuseppe Orsatti}
\affiliation{SISSA, via Bonomea 265, 34136, Trieste, Italy}
\affiliation{INFN, Sezione di Trieste}

\begin{abstract} 

\noindent
We first consider a  deterministic gas of  $N$ solitons for  the  Focusing  Nonlinear Schr\"odinger   (FNLS) equation in the limit $N\to\infty$ with a 
 point spectrum  chosen to interpolate a given spectral soliton density over a
  bounded domain of the complex  spectral plane.   
 We show that when the domain is a disk    and the soliton density  is an analytic function, then the corresponding   deterministic soliton gas surprisingly  yields  the one-soliton solution with  point spectrum the center of the disk. We call this effect {\it soliton shielding}.  We show that this behaviour  is robust and survives also for a {\it stochastic} soliton gas:  indeed, when the $N$ soliton spectrum is  chosen as random variables  either uniformly distributed on the circle, or chosen according to the  statistics of the eigenvalues of the Ginibre random matrix  the phenomenon of soliton shielding persists  in the limit $N\to \infty$.
When the domain  is an ellipse, the  soliton shielding reduces the spectral data to the soliton density concentrating between the foci of the ellipse.
The physical solution is  asymptotically step-like oscillatory, namely, the  initial profile 
is a periodic elliptic function in the negative $x$--direction while it vanishes exponentially fast in the opposite direction.
  \end{abstract}

\maketitle

\noindent
{\bf Introduction.}
The  wave propagation in a variety of physical systems  is well described by dispersive  integrable nonlinear wave equations. Integrability implies the  existence of nonlinear modes that interact elastically and are called {\it solitons}.  The inverse scattering, also called nonlinear Fourier transform, is  the tool to analyze how a general given wave packet can be viewed as a  nonlinear superposition of solitons.
Recently several investigations, both on the mathematical   and the physical side, have been carried out in which   a very large number of solitons is considered.
Coherent nonlinear superposition of many solitons occurs when one tries to optimally correlate the parameters of many nonlinear modes in order to produce a ``macroscopic'' wave profile that behaves more like a single broad wave packet than a combination of many smaller objects.  This typically occurs in small dispersion limit or semiclassical limits \cite{LL,KMM, JM14}.
Incoherent/random nonlinear superpositions of solitons are more closely related to the notion of a
soliton gas in an infinite statistical ensemble
of interacting solitons that was first introduced by Zakharov \cite{Zakharov}  for the
Korteweg de Vries  (KdV) equation.  Further  generalization  were later derived for KdV (see e.g.\cite{EL}) and for the  Focusing  Nonlinear Schr\"odinger equation (FNLS)
in \cite{EK05,EL1}. Connection between statistical properties of a soliton gas and generalized hydrodynamic has been recently established in \cite{CER21,BED22,DYC18}. Statistical properties of solutions of large set of random solitons  have been numerically investigated 
in \cite{Pelinovsky2},\cite{Pelinovsky3},\cite{Gelash19},\cite{GASR21},\cite{EL0},\cite{EL1}. Experimental realizations of 
behaviour of large sets of solitons are obtained in \cite{STBCDGPMCFER20} and \cite{RBMOM19}.
In this note,  following the lines of \cite{GGJM,GGJMM}, we consider a   soliton   gas  that originates  from the limit $N\to\infty$ of  the $N$-soliton solution of the  FNLS equation  
\begin{equation}
\label{FNLS}
i\psi_t+\frac{1}{2}\psi_{xx}+|\psi|^2\psi=0.
\end{equation}
We consider both the cases in which the $N$-soliton  spectra is chosen  in a deterministic and random way.

Let us recall the one--soliton solution, given by 
\begin{equation}
\label{one_soliton}
\psi(x,t)=2b\,\mbox{sech} [2b(x+2at-x_0)]{\rm e}^{-2i[ax+(a^2-b^2)t+\frac{\phi_0}{2}]},
\end{equation}
where $x_0$ is the initial peak position of the solition, $\phi_0$ is the initial phase, $2b$ is the modulus of the wave maximal amplitude and 
$-2a$ is the soliton velocity. The general $N$ soliton solution can be  obtained from the Zakharov-Shabat \cite{ZS80} linear spectral problem, reformulated as a {\it Riemann-Hilbert Problem}  (RHP)  for a $2\times 2$ matrix $Y^N(z;x,t)$ 
with the following data \cite{Faddev}:   the discrete spectrum $S:=\{z_0;\dots;z_{N-1};\bar{z}_0;\dots;\bar{z}_{N-1}\}$, $z_j\in\mathbb{C}^+$ the upper half space,  and its norming constants $\{c_0,\dots,c_{N-1}\}$ with $c_j\in\C$.  Here and below $\bar{z}$ stands for the complex conjugate of $z$.


 The matrix $Y^N(z;x,t)$  is analytic for 
 $z\in \C\backslash S$ and has   {\it simple } poles in $S$   with the  residue condition
 \begin{equation}
 \label{eq:Res_cond}
  \begin{split}
    \res\limits_{z=z_j}Y^N(z) &= \lim_{z \to z_j} Y^N(z)
        \begin{pmatrix}
          0 & 0\\
          c_j {\rm e}^{2\theta(z,x,t)} & 0\\
        \end{pmatrix}\\
         \res\limits_{z=\overline{z}_j}Y^N(z) &= \lim_{z \to \bar{z}_j} Y^N(z)
        \begin{pmatrix}
          0 & -\bar{c}_j {\rm e}
          ^{-2\theta(z,x,t)}\\
          0 & 0\\
        \end{pmatrix}\\
        Y^N(z)&= \mathbb{I}+\mathcal{O}\left(\frac{1}{z}\right),\;\;\mbox{as $z\to\infty$},\\
  \end{split}
\end{equation}
where  $\theta(z,x,t)= i(z^{2}t + zx)$  and $\mathbb{I}$  is the identity matrix.  The   equations \eqref{eq:Res_cond}  uniquely determine $Y^N(z;x,t)$ as a rational matrix function of $z$ in the form
\begin{equation}
\label{solution}
Y^N(z;x,t)= \mathbb{I}+\sum_{j=0}^{N-1}\dfrac{\begin{pmatrix}f_j(x,t)&0\\
g_j(x,t)&0\end{pmatrix}}{z-z_j}+\sum_{j=0}^{N-1}\dfrac{\begin{pmatrix}0&-\overline{g_j(x,t)}\\
0&\overline{f_j(x,t)}\end{pmatrix}}{z-\overline{z_j}}\,,
\end{equation}
where the coefficients $f_j(x,t)$ and $g_j(x,t)$ are determined  from a linear system by imposing the residue conditions in \eqref{eq:Res_cond}.
The solution of the FNLS equation is recovered from $Y^N(z;x,t)$ by the relation
\begin{equation}
\label{Nsoliton}
\psi_N(x,t)=2i \lim_{z \to \infty}z(Y^N(z;x,t))_{12}
\end{equation}
which gives the $N$-soliton solution in the form $\psi_N(x,t)=-2i\sum_{j=0}^{N-1}\overline{g_j(x,t)}$.
In the case of one soliton solution, we have  that the point spectrum $z_0=a+ib$  determines the speed and amplitude of the soliton  \eqref{one_soliton} and  the coefficient $c_0$ determines the position  $x_0 =\frac{\ln(|c_0|)}{2b}$  of the soliton peak  and the phase $\phi_0=\frac{\pi}{2}+\arg(c_0)$ of the soliton. 

The  FNLS equation can have  a soliton  of order $N$  when the  matrix function $Y^N(z)$  has  a pole of order $N$. Such a solution can be viewed as $N$  soliton solution where the simple poles coalesce to a pole of order $N$. The limit as $N\to\infty$ of such solution has been studied in \cite{Bilman1, Bilman2} where it has been shown that
its  near field structure is described by  the Painlev\'e III equation. An analogous asymptotical study has been performed for breathers in \cite{Bilman3}.

In this letter we consider the case when the  norming constants  $\{c_j\}_{j=0}^{N-1}$, scale as $1/N$ as  the number $N$  of simple poles (i.e. the number of solitons)  tends to infinity.
On the physical side, scaling the norming constants to be small means that the individual solitons are centered at positions that are logarithmically large in $N$,  so that in the finite part of the $(x,t)$  plane only  the tails  of the solitons add up.  The resulting gas of  solitons  is a condensate in the terminology of \cite{EL1}.

Differently from \cite{GGJM}, \cite{GGJMM} where the  infinite set of solitons is obtained by letting 
the soliton spectra accumulate on lines of the spectral complex plane, here we consider the case in which  soliton spectra  accumulate  on  one or more  simply connected  bounded domains $\mathcal{D}$ of the complex  upper  plane  $\mathbb{C}^+$ and their complex conjugate $\overline{\mathcal{D}}$.
We let the number of solitons goes to infinity in such a way that their point spectrum  $z_j$ ($\bar{z}_j$)  fills {\bf uniformly} the domain $\mathcal{D}$.
The corresponding norming constants $c_j$ are interpolated by a smooth function $\beta(z,\bar{z})$, namely
 \begin{equation}
 \label{cj}
 c_{j}=\frac{\mathcal{A}}{\pi N} \beta(z_j,\bar{z}_j),
\end{equation}
where  $\mathcal{A}$ is  the area of the  domain $\mathcal{D}$ and $N$ is the total number of solitons.
%

The remarkable emerging feature   is that  as $N\to\infty$, for certain types of domains and densities, we have a ``soliton shielding'', namely, the gas behaves as a {\it finite} number of solitons. This happens for example if the distribution function is $\beta(z,\ov z)=\overline{z}^{n-1}r(z)$ with $r(z)$ an  analytic function in $\mathcal{D}$,   and  the domain  is described by  $\mathcal{D}:= \{ z\in \mathbb{C} \text{ s.t. }  |(z-d_0)^n-d_1|<\rho\},\;\;n\in\mathbb{N},$ with $d_0\in\mathbb{C}^+$,   $|d_1|$  and $ \rho>0$ sufficiently small so that $\D\in\mathbb{C}^+$. Then   the   deterministic soliton gas is equivalent to a $n$-soliton solution.  In the case $n=1$, the domain 
$\mathcal{D}$ is  a disk  centered at $\lambda_0=d_0+d_1$  and the infinite number of solitons superimpose nonlinearly in their tails to produce a single soliton  solution with point spectrum      $\lambda_0$ and  norming constant   equal to $\rho^{2}r(\lambda_0)$. We are going to see that this behaviour persists also when the   $N$ soliton spectrum   is a random variable distributed according to the Ginibre ensemble \cite{Ginibre} or the uniform distribution on the disk. 


When the  domain $\mathcal{D}$ is an ellipse    we show that such  deterministic  soliton gas is a step-like periodic elliptic  wave  at $x=-\infty$ and rapidly decreasing at $x=+\infty$  as in  \cite{GGJM}.


\vskip 0.1cm
\noindent
{\bf Deterministic soliton gas.}  In order to obtain the limit of the $N$-soliton solution as $N\to \infty$, we impose that the norming constants $c_j$  scale as  $1/N$.   
%
%
Then we use  a  transformation that removes the singularities of $Y^N$. Indeed   let $\gamma_+$ be a closed anticlockwise oriented  contour  that encircles all the poles in the upper 
half space  and $D_{\gamma_+}$ the finite domain with boundary $\gamma_+$ and similarly we define    $\gamma_{-}=-\overline{\gamma_+}$   and  $D_{\gamma_-}$
encircles all the poles in the lower half space.
%

 One  ends up with the  RHP for the matrix function $\widetilde{Y}^N(z;x,t)$  analytic in $\mathbb{C}\backslash\{\gamma_+\cup\gamma_-\}$, subject to the conditions
 \begin{equation}
  \label{eq:R-H_sol}
  \begin{split}
  &{\widetilde{Y}_+^N(z,x,t)=\widetilde{Y}_-}^N(z,x,t)\tilde{J}_N(z,x,t),\;\;\;z\in \gamma_+\cup\gamma_-\\
  & \widetilde{Y}^N(z;x,t)= \mathbb{I}+\mathcal{O}\left(\frac{1}{z}\right),\;\;\mbox{as $z\to\infty$},
  \end{split}
  \end{equation}
  where  the subscripted  {$Y_\pm$  denote the left/right boundary values   along the oriented contour} and 
  \begin{equation}
  \label{Jt}
  \tilde{J}_N(z,x,t)=\left\{
\begin{array}{ll}
 \begin{pmatrix}
    1 & 0\\
   -\tiny{ \sum\limits_{j=0}^{N-1} \frac{c_{j}{\rm e}^{2\theta(z_j,x,t)}}{z-z_{j}}}& 1
  \end{pmatrix}, &z\in\C_+\\
  \begin{pmatrix}
    1 &    \tiny{\sum\limits_{j=0}^{N-1} \frac{\bar{c}_{j}{\rm e}^{-2\theta(\ov z_j,x,t)}}{z-\bar{z}_{j}}}\\
0& 1
  \end{pmatrix},&z\in\C_-\,.
\end{array}
\right.
\end{equation}
We call the matrix $\tilde{J}_N(z,x,t)$ the {\it jump matrix}.
The solution $\widetilde{Y}^N(z,x,t)$ is obtained from $Y^N(z,x,t)$ by the relation 
$\widetilde{Y}^N(z,x,t)=Y(z,x,t)$ for $z$ in $\mathbb{C}\backslash\{  D_{\gamma_+}\cup D_{\gamma_-}\}$
and $\widetilde{Y}^N(z,x,t)=Y(z,x,t)\tilde{J}_N(z,x,t)$ for $z\in D_{\gamma_+}\cup D_{\gamma_-}$.  In this case  the coefficients $f_j$ and $g_j$ in \eqref{solution}  are recovered by imposing  $\widetilde{Y}^N(z,x,t)$ to be analytic at $z_j$ and $\overline{z_j}$ for $j=0,\dots, N-1$.

Let $\mathcal{D}$ be a  domain  so  that   the closure of $\mathcal{D}$  is  strictly contained in the domain $D_{\gamma_+}$  bounded by  $\gamma_+$ and the closure of $\overline{\mathcal{D}}$ 
is completely contained in the domain  $D_{\gamma_-}$ bounded by  $\gamma_-$. We let the number of solitons goes to infinity in such a way that their point spectrum  $z_j$ ($\bar{z}_j$)  fills {\bf uniformly} the domain $\mathcal{D}$  contained in $\gamma_+$ and we choose the norming constants $c_j$ as in \eqref{cj} so that
$$
\sum_{j=0}^{N-1}\frac{c_j}{(z-z_j)}=\sum_{j=0}^{N-1}\frac{\mathcal{A}}{\pi N} \frac{\beta(z_j,\bar{z}_j)}{z-z_j} \mathop{\longrightarrow}_{N\to\infty} \iint_{\mathcal{D}}\frac{\beta(w,\bar{w})}{z-w}\frac{\d^2w}{\pi},
$$
%
where  the infinitesimal area measure is  $\d^2w=(\d\overline{w}\wedge \d w)/(2i)$.
Consequently  the RH-problem  \eqref{eq:R-H_sol} becomes
\begin{equation}
  \label{eq:R-H_sol2}
  \begin{split}
  &\widetilde{Y}_+^\infty(z,x,t)=\widetilde{Y}_-^\infty(z,x,t)\tilde{J}_\infty(z,x,t),\;\;\;\tilde{J}_\infty(z,x,t)=\\
  & \begin{pmatrix}
    1 & \iint\limits_{\overline{\mathcal{D}}}\frac{{\rm e}^{-2\theta(w,x,t)} \beta^{*}(w,\bar{w})\d^2w}{\pi(z-w)}\chi_{\gamma_{-}}\\
    \iint\limits_{\mathcal{D}}\frac{{\rm e}^{2\theta(w,x,t)} \beta(w,\bar{w})\d^2w}{\pi(w-z)}\chi_{\gamma_{+}} & 1
  \end{pmatrix}
\\
   & \widetilde{Y}^\infty(z;x,t)= \mathbb{I}+\mathcal{O}\left(\frac{1}{z}\right),\;\;\mbox{as $z\to\infty$},
\end{split}
\end{equation}
with $\beta^{*}(w,\bar{w})= \overline{\beta(\bar{w},w)}$. The limiting FNLS solution is given by 
\begin{equation}
\label{infinitysol}
\psi_{\infty}(x,t)=2i \lim_{z \to \infty}z( \widetilde{Y}^\infty(z;x,t))_{12}.
\end{equation}
For a general bounded domain $\mathcal{D}$ and smooth function $\beta(z,\overline{z})$, the class of solutions of FNSL obtained from \eqref{eq:R-H_sol2} and \eqref{infinitysol}  is  unexplored.
 In the case   $\beta(z,\bar{z})= n\overline{z}^{n-1}r(z)$,    with  $r(z)$   analytic in $\mathcal{D}$, we can apply Green theorem for $z\notin \mathcal{D}$
 and obtain
 \begin{equation}
 \label{eq:int_prob1}
  \iint\limits_{\mathcal{D}}\frac{{\rm e}^{2\theta(w,x,t)} \beta(w,\bar{w})\d^2w}{\pi(z-w)}=
   \int_{\partial\mathcal{D}}\frac{r(w)\overline{w}^{n}{\rm e}^{2\theta(w;x,t)}}{ z-w} \frac{\d w}{2\pi  i },
   \end{equation}
   and similarly for the integral over $\overline{\mathcal{D}}$.

 For sufficiently smooth simply connected domains $\mathcal D$ the  boundary $\partial \mathcal{D}$  can be  described by the so--called   Schwarz function  $S(z)$ \cite{Gustafsson}  of the domain $\mathcal{D}$  through the equation
\[
\overline{z}=S(z).
\]
 The Schwarz function admits analytic extension to  a maximal domain $\mathcal D^0\subset \mathcal D$. For example,  for quadrature domains,  $\mathcal D^0$ is just $\mathcal D$ minus a finite collection of points  \cite{Gustafsson}. The simplest such quadrature domain is the disk, which is one of our examples below. 
For  other classes of domains we have that $\mathcal D\setminus \mathcal D^0$  may consist of a {\it mother-body}, i.e.,  a collection of smooth arcs \cite{AHM}. An example of this is the ellipse, which will be our second example.\\[8pt] 
\noindent{\bf Shielding of soliton gas for  quadrature domains.}\\[1pt]
We start by considering the   class of domains  
\begin{equation}
\label{D}
\mathcal{D}:= \left \{ z\in \mathbb{C} \text{ s.t. }  \Big|(z-d_0  )^m-d_{1}\Big|<\rho\right\},\;\;m\in\mathbb{N},
\end{equation}
with $d_0\in\mathbb{C}^+$ and  $|d_1|, \rho>0$ sufficiently small so that $\D\in\mathbb{C}^+$. When $m=1$ such domain coincides with the disk $\mathbb{D}_\rho(\lambda_0)$ of radius $\rho>0$
centred at  $\lambda_0= d_0+d_1$.
When $m>1$ the domain $\D$ has a $m$--fold symmetry about $d_0$ and  is simply connected if $|d_1|\leq \rho$, and otherwise it has $m$ connected components \cite{BGM}.
The boundary of $\mathcal{D}$  is described by 
\begin{equation}
\label{Schwartz1}
\overline{z}=S(z),\quad S(z)=\overline{d_0}+\left(\overline{d}_1+\frac{\rho^2}{(z-d_0)^m-d_1}\right)^{\frac{1}{m}}.
\end{equation}
 {\bf   The $n$-soliton solution.} This  solution  is obtained  from \eqref{eq:int_prob1} by choosing   $m=n$ in \eqref{Schwartz1}.   
 We then substitute $\ov w =S(w)$ in the contour integral \eqref{eq:int_prob1} and use the residue theorem at the  $n$  poles given by the solution $\{\lambda_0,\dots, \lambda_{n-1}\}$ of the equation $(z-d_0)^n=d_1$.  Then 
 
 \begin{align*}
  &\int_{\partial\mathcal{D}}\frac{\overline{w}^n\,r(w){\rm e}^{2\theta(w;x,t)}}{ z-w} \frac{\d w}{2\pi i }=
  \int_{\partial \mathcal{D} }S(w)^nr(w)\dfrac{{\rm e}^{2\theta(w;x,t)}}{z- w}\frac{\d w}{2\pi i }\\
  &= \rho^2\sum_{j=0}^{n-1}\frac{r(\lambda_j)}{\prod_{k\ne j}(\lambda_j - \lambda_k)}\dfrac{{\rm e}^{2\theta(\lambda_j;x,t)}}{z-\lambda_j},\quad z\notin \mathcal{D},
 \end{align*}
 which gives, up to a sign the entry $21$ of the jump matrix \eqref{Jt}.
 Namely the solution $\psi_{\infty}(x,t)$ in \eqref{infinitysol} coincides with the $n$ soliton solution  $\psi_n(x,t)$ in \eqref{Nsoliton} with spectrum $\{\lambda_0,\dots,\lambda_{n-1}\}$ and corresponding norming constants
 $c_j=\rho^2 r(\lambda_j)/\prod_{k\ne j}(\lambda_j - \lambda_k)$ for $j=0,\dots, n-1$.\\
 {\bf One soliton solution.}
 In particular, in the case $n=m=1$ and  $\D=\mathbb{D}_\rho(\lambda_0)$ the disk centred at $\lambda_0 = d_0+d_1$ of radius $\rho$,
 we obtain  exactly the RH-problem \eqref{eq:R-H_sol}  for $N=1$ and $c_0=\rho^2r(\lambda_0)$.
Namely we recover  the one soliton  solution \eqref{one_soliton}  of the FNLS~\eqref{FNLS}  equation
with   $\lambda_{0}= d_0+d_1=a + i b$, with peak  position $x_0$ and  phase shift $\phi_0$ given  respectively  by 
\begin{equation}
  \label{eq:soliton_shift}
  x_0 := \frac{\log(|\rho^2 r(\lambda_{0})|) - \log(2b)}{2b}, \; \phi_0:= \arg(r(\lambda_0))-\frac{\pi}{2}.
  \end{equation}
  We observe that the radius $\rho$ of the disk   and the value of the function $r(z)$ at $\lambda_0$ contribute to the phase shift of the soliton
  but not to its amplitude or velocity, which are uniquely determined  by the center of the disk $\lambda_0$.\\[5pt]
\noindent
{\bf Soliton solution of order $n$.} By considering $m=1$, namely, the disk $\mathbb{D}_\rho(\lambda_0)$ and  $\beta(z)=n(\bar{z}-\bar{\lambda}_0)^{n-1} r(z)$  for $n>1$  one obtains the soliton solution of order $n$.
This degenerate solution  and the limit $n\to\infty$  has been extensively analyzed in \cite{Bilman2}.

\begin{figure}
\includegraphics[width=0.2
\textwidth]{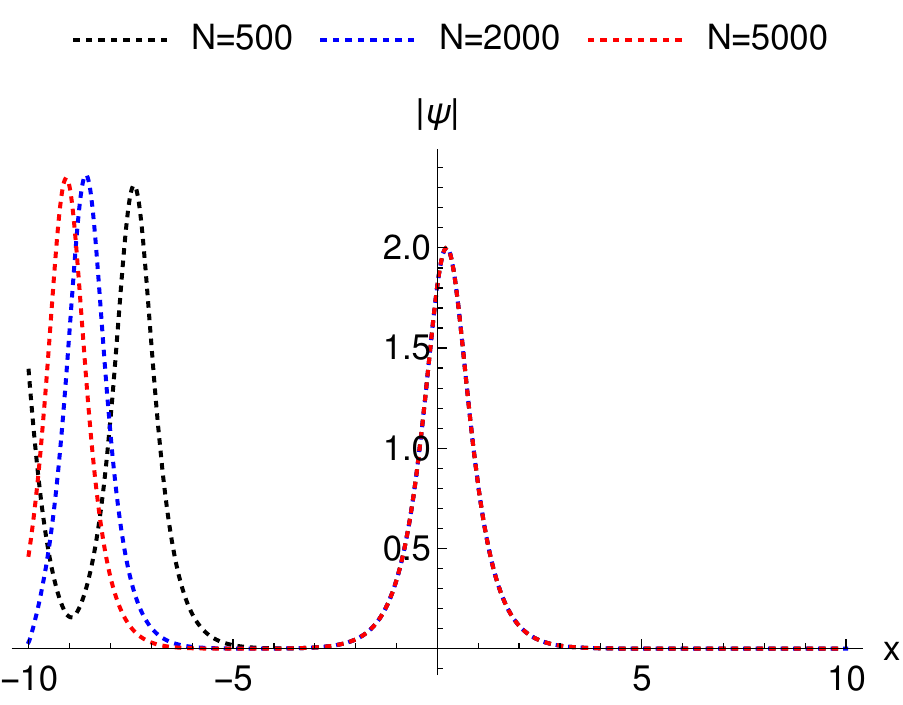}
\includegraphics[width=0.2
\textwidth]{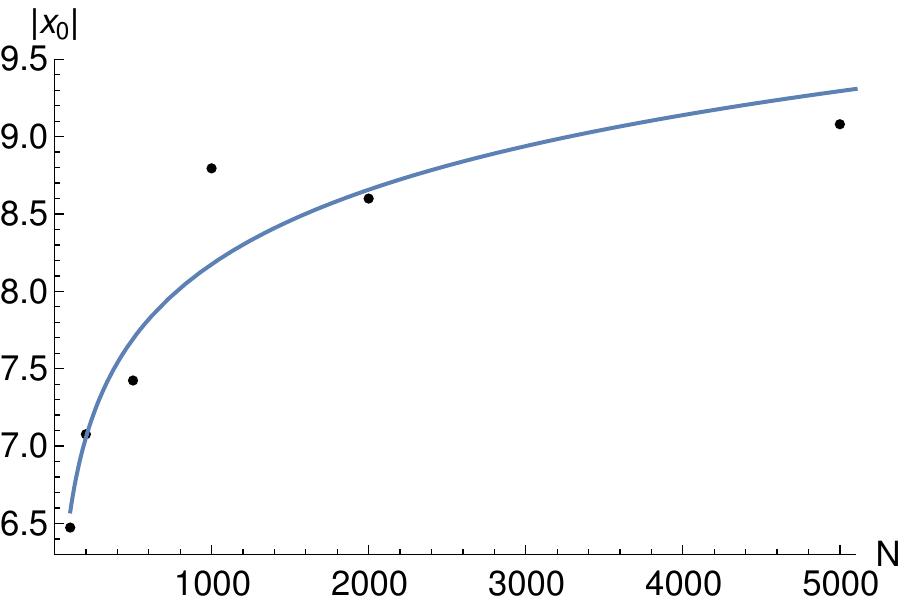}
\caption{
On the left, the plot of the gas that approximates the area measure using $N$ Fekete points for $N=500,2000,5000$, all centred in a disk of ray $1/10$ and center $\lambda_0=i$ (with $\beta(z)=\pi/\rho^2$), and the emerging limiting (one-soliton) solution $\psi^{\infty}(x,t)$ centred at $x=0.226$. On the right,  a fit (with a curve of the form $q + p\log(N)$) of the distance between the peak of the  limiting soliton solution  $\psi^{\infty}(x,t)$  and the first peak of the remaining part of the solution that is going to infinity as $N\to\infty$. }
\label{plot}
\end{figure}

\noindent
{\bf Remark}.
In Figure \ref{plot} we plot the resulting ``effective'' soliton using an approximation of the uniform measure on the unit disk by means of $N$ Fekete points, namely the set of $N$ points described by the vector $\boldsymbol{w}=(w_0,\dots,w_{N-1})$ that minimizes the energy
\begin{equation}
\label{Energy}
E(\boldsymbol{w})=
-2\sum_{0\leq j< k\leq N-1} \log|w_j-w_k|+\frac{N}{2}\sum_{j=0}^{N-1} |w_j|^2,
\end{equation} 
(suitably translated/rescaled)
over all possible configurations. Then  the uniform measure on the disk $D_\rho(\lambda_0)$ is obtained by the rescaling $z_j=\rho(w_j-\lambda_0)$.
 The train of solitons on the left (albeit slowly) will move towards $-\infty$ as $\mathcal O(\log N)$.

\vskip 5pt
\noindent
{\bf Elliptic Domain.}
We now consider the  case in which $\D$ coincide with an elliptic domain   $\mathcal{E}$ and with $\beta (z,\bar{z})=r(z)$ analytic.
For the sake of simplicity, we  assume  that the focal points   $i \alpha_{1}$ and $i \alpha_2$ of the ellipse lie  on  the imaginary axis and  $\alpha_2>\alpha_1>0$. The equation of the ellipse is
$\sqrt{x^2+(y-\alpha_1)^2}+\sqrt{x^2+(y-\alpha_2)^2}=2\rho>0$, where $\rho$ is chosen sufficiently small so that $\mathcal{E}$ lies in the upper half space.
 %
%
We choose  $\beta(z)=r(z)$  to be analytic in $\mathcal{E}$.
In this case  in equations \eqref{eq:int_prob1} $n=1$ and one has to consider the Schwarz function of the ellipse, namely $ \bar{z}=S(z):$ 
\begin{equation}
\begin{split}
   S(z)&=\left(1 -\frac{2\rho^2}{c^2}\right)(z-i y_0)+2\frac{\rho}{c^2}\sqrt{\rho^2-c^2}R(z) -iy_0,\\
\end{split}
\end{equation}
 where  $R(z):= \sqrt{ (z-i\alpha_1)(z-i\alpha_2)}$,  $ y_0=\frac {\alpha_1+\alpha_2}2$ and   $c = \frac{\alpha_2-\alpha_1}2$.
The function $S(z)$ is analytic  in $\C$  away   from  the segment  $\mathcal{I}:=[i\alpha_1,i\alpha_2]$, with boundary values $S_{\pm}(z)$. For $z\notin \mathcal{E}\cup\overline{\mathcal{E}}$, 
the integral along  the boundary $\partial \mathcal{E}$   ($\partial \overline{ \mathcal{E}})$ of the ellipse in \eqref{eq:int_prob1}  can be deformed to a line integral  on  the segment $\mathcal{I}=[i\alpha_1,i\alpha_2]$  ($\overline{\mathcal{I}}:=[-i\alpha_2,-i\alpha_1]$), namely
\[
 \int_{\partial\mathcal{E}}\frac{r(w)\overline{w}{\rm e}^{2\theta(w;x,t)}}{ z-w} \frac{\d w}{2\pi  i }=\int_{\mathcal{I}}\frac{r(w)\delta S(w){\rm e}^{2\theta(w;x,t)}}{ z-w} \frac{\d w}{2\pi  i },
\]
where $\delta S(z)= S_{+}(z) - S_{-}(z)$.
Next we define 
\begin{equation}
\Gamma(z):=\left\{
\begin{array}{ll}
&\widetilde{Y}^\infty(z),\quad z\in\mathbb{C}\backslash\{D_{\gamma_+}\cup\mathcal{D}_{\gamma_-}\}\\
&\widetilde{Y}^\infty(z)J(z),\quad z\in D_{\gamma_+}\cup\mathcal{D}_{\gamma_-}
\end{array}
\right.
\end{equation}
where  $\;\; J(z)=$
\begin{equation*}
  \begin{split}
 \begin{pmatrix}
\hspace{-20pt}    1 &\hspace{-60pt}\displaystyle \int\limits_{\overline{\mathcal{I}}}\frac{r^*(w)\delta S^*(w){\rm e}^{-2\theta(w;x,t)}}{ w-z} \frac{\d w}{2\pi  i }\chi_{\mathcal{D_{\gamma_-}}} \\
\displaystyle   \int\limits_{\mathcal{I}}\frac{r(w)\delta S(w){\rm e}^{2\theta(w;x,t)}}{ z-w} \frac{\d w}{2\pi  i }\chi_{\mathcal{D_{\gamma_+}}} & 1
  \end{pmatrix}.
  \end{split}
\end{equation*}
In this way $\Gamma(z)$ does not have a jump on $\gamma_+\cup\gamma_-$. Since $J(z)$ has a jump in $\I\cup\overline{\I}$
  it follows that  $\Gamma(z)$ is  analytic in $\mathbb{C}\backslash\{\I\cup\overline{\I}\}$  with jump conditions
\begin{equation}
\begin{split}
  \label{eq:ell_R-H-prob}
    \Gamma_{+}(z)&=\Gamma_{-}(z){\rm e}^{\theta(z;x,t) \sigma_{3}}G(z){\rm e}^{-\theta(z;x,t) \sigma_{3}}\\
    G(z)&= \begin{pmatrix}
    1 &\chi_{\overline{\I}} \delta S^*(z) r^*(z) \\
    -\chi_{\I}\delta S(z) r(z) & 1
  \end{pmatrix},
\end{split}
\end{equation}
 and
$\Gamma(z)=\mathbb{I}+O(\frac{1}{z})$, as $z\to\infty$.
We can find the same RHP~\eqref{eq:ell_R-H-prob} also when we study the problem~\eqref{eq:R-H_sol} with an infinite number of spectral points uniformly distributed along the segments  $\I\cup\overline{\I}$.

For $t=0$, the initial datum $\psi_0(x)$  associated to the  solution of  the RHP~\eqref{eq:ell_R-H-prob}  turns out to be step-like  oscillatory. Indeed from the steepest descent  method following the lines \cite{GGJM} as $x\to-\infty$ we derive \cite{InProgress} the elliptic function
\begin{equation}
       \label{eq:NLS-t}
   \psi_{0}(x)=i (\alpha_2 +\alpha_1) 
		\mbox{dn} \left[ (\alpha_2 +\alpha_1)  ( x -x_0); m \right]+\mathcal{O}(x^{-1}),
\end{equation}
where $\mbox{dn}(z;m)$ is the Jacobi elliptic function of modulus $m=\frac{4 \alpha_2 \alpha_1}{(\alpha_2+ \alpha_1)^2}$  and $x_0$ 
 is a constant which depends on $\delta S(z)$,  $ \beta(z)$ and the geometry of the problem.
For   $x \to+\infty$ the initial datum  goes to zero exponentially fast.
 When $\alpha_1\to\alpha_2$  the ellipse degenerates to the circle and one  recovers the one soliton solution.\\[3pt]
\noindent{\bf Random soliton gas with  Ginibre and uniform statistics}\\[1pt]
Let us now introduce randomness in the system by choosing the points $ z_j=\rho(w_j-\lambda_0)\chi_{\C^+}$ with $(w_0,\dots, w_{N-1})\in\C^N$  distributed according to the probability density  ({\it Ginibre ensemble})
\begin{equation}
\label{muN}
\mu_N=\frac{1}{Z_N}e^{-E(w_0,\dots, w_{N-1})}d^2w_0\dots d^2w_{N-1},
\end{equation}
where $Z_N$ is the normalizing constant and $E(w_0,\dots, w_{N-1})$ is the energy defined in \eqref{Energy}.
In the limit $N\to\infty$  the random  points $ \{w_0,\dots, w_{N-1}\}$  fill uniformly  the  unit  disk  centered at zero  ( see e.g. \cite{Ginibre}).   
For any smooth function $h:\C \to \C$, let us consider the random variable  $X^N_h:= \sum_{j=1}^N h\left(w_j\right)$. It is known \cite{Rider_Virag} that  
\be
\frac 1N\mathbb E[ X_h]\mathop{\longrightarrow}\limits_{N\to \infty} \int_{|w|\leq 1} h(w) \d^2 w,
\ee
where $\mathbb{E}$ is the expectation with respect to the probability measure $\mu_N$.
Actually more is true \cite{AHM} \cite{Rider_Virag}: the limit of the random variable $X_h-\mathbb E[ X_h]$  converges to a normal random variable $\mathcal{N}(0,\sigma)$ centred at zero and  with finite variance $\sigma^2$ depending on $h$. 

\begin{figure}
\includegraphics[width=0.22
\textwidth]{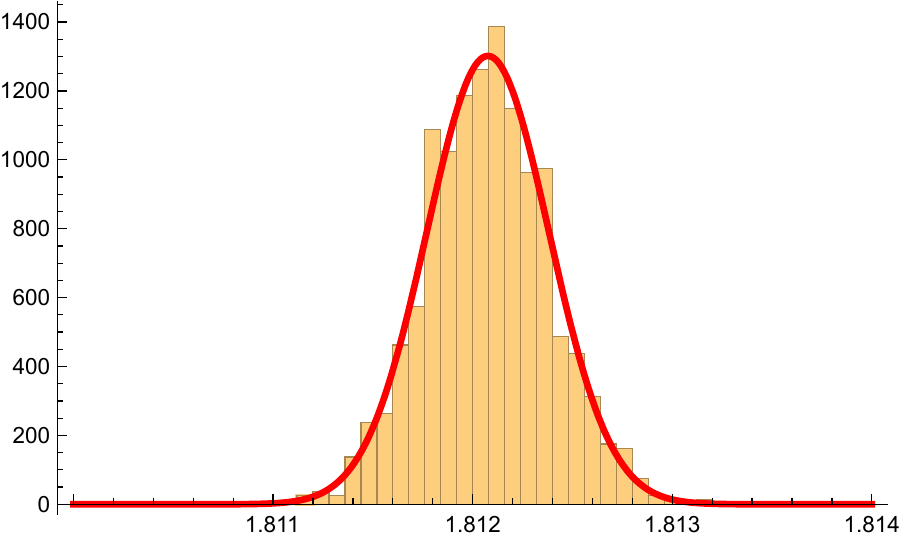}
\includegraphics[width=0.22
\textwidth]{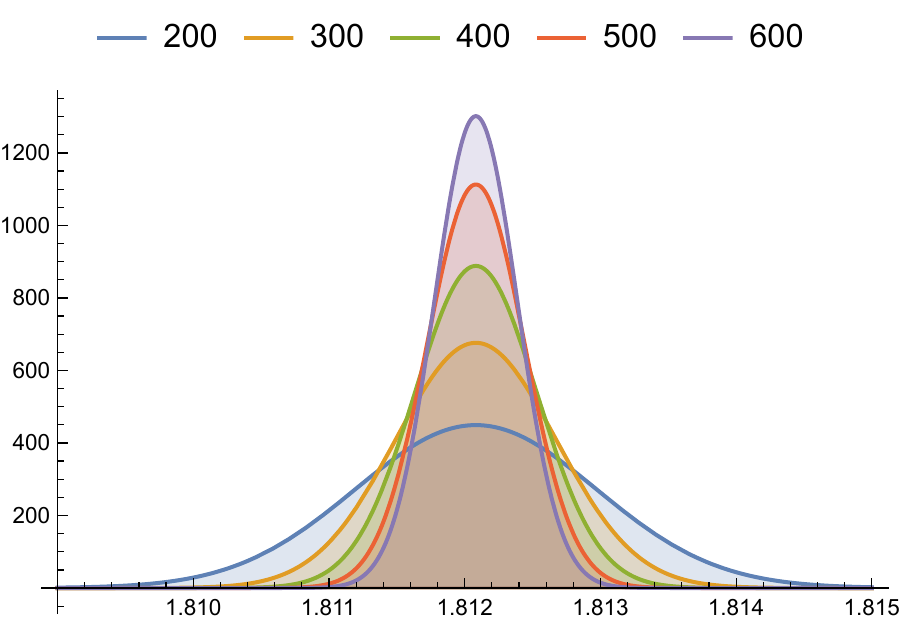}
\includegraphics[width=0.22
\textwidth]{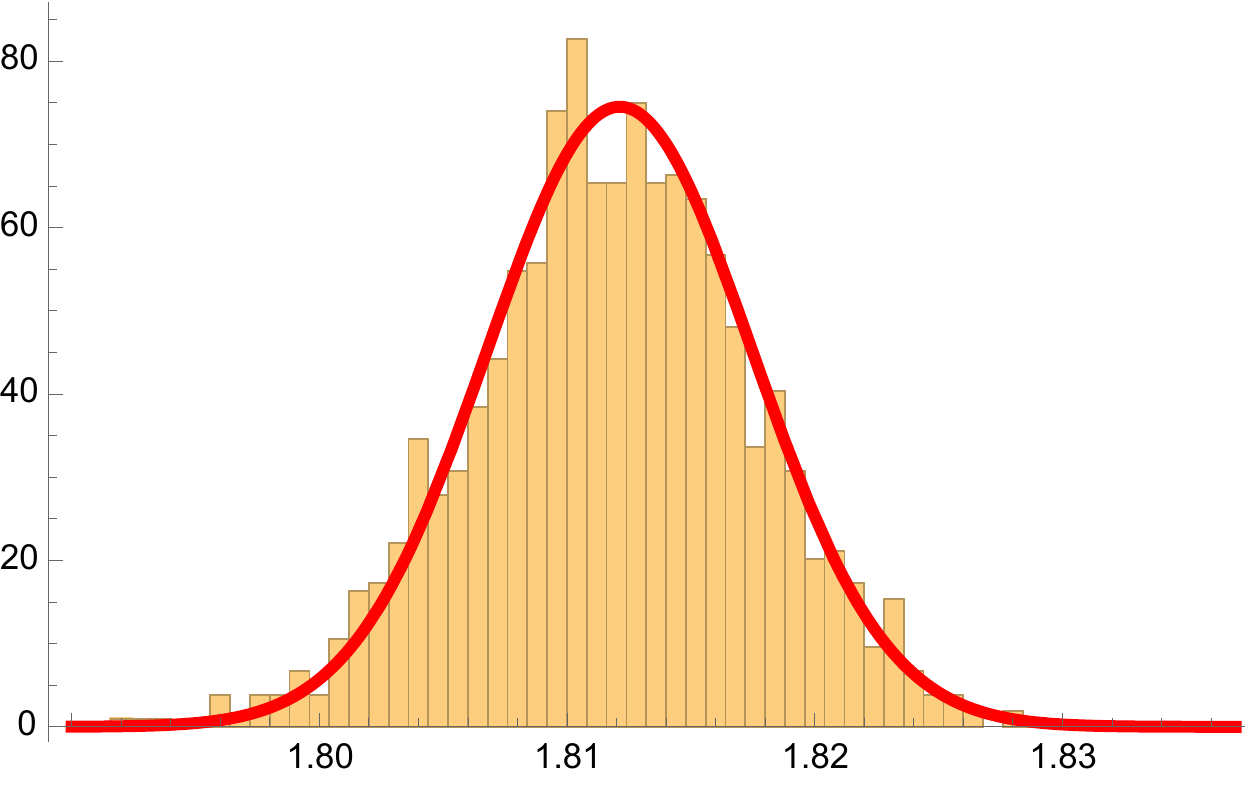}
\includegraphics[width=0.22
\textwidth]{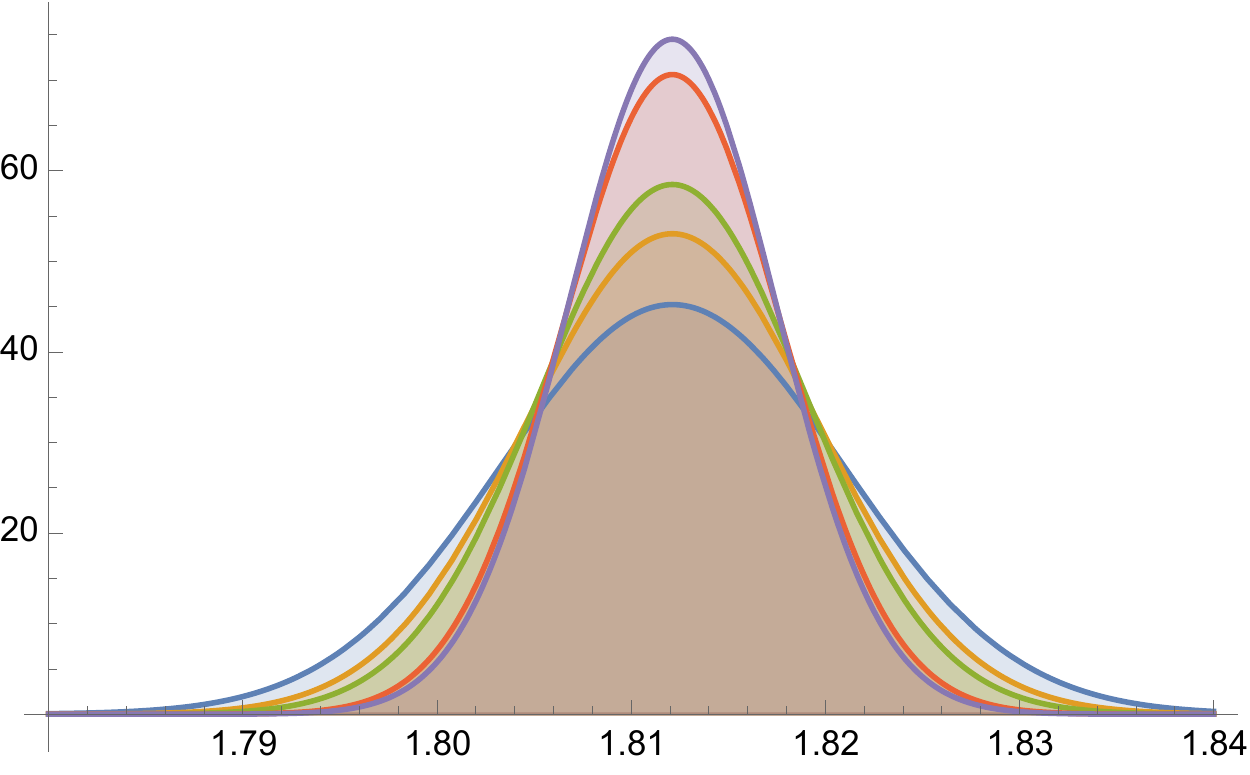}
\caption{ Left: the Gaussian fitting  of the fluctuations of the $N=600$ soliton  solution $\psi_N(0,0)$  with respect to the limiting solution $\psi^{\infty}(0,0)\simeq1.812$   and $1000$ trials. The point spectrum is sampled in the disk $\mathbb{D}_{1/10}(i)$ according to the Ginibre ensemble (top) and the uniform distribution (bottom). On the right figure the corresponding  Gaussian fitting for $N=200,300,400,500,600$. The Gaussian distribution is centred at  $\psi^{\infty}(0,0)$ and the error $\sigma$  scales  numerically
 as $0.178/N $ (Ginibre) and $0.129/N^{\frac{1}{2}} $  (uniform distribution). The scaling does not depend on the point $x=0, t=0$ chosen to make the statistics. }
\label{plot2}
\end{figure}

From the above arguments  it is expected that  the jump of the RH-problem \eqref{eq:R-H_sol2}, in probability, satisfies  $$\mathbb{P}\left(\left|\sum_{j=0}^{N-1}\frac{\mathcal{A}}{ N} \frac{\beta(z_j,\bar{z}_j)}{z-z_j}- \iint_{\mathcal{D}}\frac{\beta(w,\bar{w})}{z-w}\d^2w\right|>\epsilon\right)=\mathcal{O}(\frac{1}{N}),$$
for  $z\notin\mathcal{D}$.
Using   small norm arguments  on  the  RH-problem  \cite{Zhou}, one may argue   that the  random $N$ soliton solution
 $\psi_{N}(x,t,z_0,\dots, z_{N-1})$  converges  as $N\to\infty$  in probability to the one-soliton solution  $\psi_{\infty}(x,t)$. 
 Similar arguments can be used  also when the soliton spectrum is sampled according to the uniform distribution  on the unit disk.  The complete mathematical  proof would require a more elaborated argument, which is  postponed to a subsequent publication.
 From numerical simulations, the fluctuations  of $\psi_{N}(x,t,z_0,\dots, z_{N-1})$ around the limiting value $\psi_{\infty}(x,t)$  are Gaussian with error that decreases at the rate  $\mathcal O(N^{-1})$, when the  random points $\{z_0,\dots,z_{N-1}\}$  are sampled  from  the Ginibre ensemble while the rate is $\mathcal O(N^{-1/2})$ for the uniform distribution on the disk, see Figure~\ref{plot2}.

 \noindent
{\bf Conclusions.}
We have considered a  gas of  $N$ solitons solution of  the  FNLS equation in the limit $N\to\infty$. The soliton spectrum $\{z_j\}_{j=0}^{N-1}$  is chosen at first as the discretization of the uniform measure of a  compact domain $\mathcal{D}$  of the complex upper half space and the norming constants
$\{c_j\}_{j=0}^{N-1}$ are interpolated by a smooth function  $\beta(z,\overline{z})$,  namely $  c_{j}=\frac{\mathcal{A}}{\pi N} \beta(z_j,\bar{z}_j)$ where $\mathcal{A}$ is the area 
measure of the domain $\mathcal{D}$.  We  then  showed that when the  domain  $\D$  is a disk    and the soliton density $\beta(z,\overline{z})$  is an analytic function, then the corresponding   $N$-soliton solution condensates   in the limit $N\to\infty$ and fixed 
$(x,t)$,  to the one-soliton solution with point spectrum  coinciding with  the center of the disk.  We call this surprising effect {\it soliton shielding} because the interaction of  infinite  solitons 
  reduces out  to a one-soliton solution. 
  Our  result is robust and persists also when the soliton spectrum is a random variable sampled  according to the Ginibre ensemble or the uniform distribution on the disk.

  The determination of the $N$-soliton solution in the double scaling limit $N\to\infty$ and $x\to\infty$ in such a way that  $x\simeq \log N$  remains a challenging open problem.
 
  For other choices of domains $\D$ or  density  $\beta(z,\overline{z})$ we  obtained  a $n$-soliton solution or a one-soliton solution of order $n$. When the domain  $\D$   is an ellipse, we showed that  the spectral measure concentrates on lines connecting the foci of the ellipse and  the soliton gas initial datum   is asymptotically step-like oscillatory.
%
\color{black}
\begin{acknowledgments}
\noindent{\bf Acknowledgments.} We are grateful to K. Mc~Laughlin for useful discussions and the {\em Isaac Newton Institute for Mathematical Sciences} for support and hospitality during the program ``Dispersive hydrodynamics: mathematics, simulations and experiments, with application in nonlinear waves'', EPSRC Grant Number EP/R014604/1.
T.G. and G.O. acknowledge the support from H2020  grant No. 778010 {\em  IPaDEGAN},  the support of INdAM/GNFM and the  research project Mathematical Methods in Non Linear Physics (MMNLP), Gruppo 4-Fisica Teorica of INFN. 

%
\end{acknowledgments}


\end{document}